\documentclass[%
 reprint,
 amsmath,amssymb,
 aps,
prb,
superscriptaddress
]{revtex4-2}

\usepackage{svg}
\usepackage{graphicx}
\usepackage{dcolumn}
\usepackage{bm}
\usepackage{physics}
\usepackage{siunitx}
\usepackage{upgreek}
\usepackage{hyperref}
\usepackage[capitalise]{cleveref}
\usepackage{mathtools}
\usepackage{color}
\usepackage{soul}

\usepackage[textsize=small]{todonotes}
\usepackage[deletedmarkup=xout]{changes}
\definechangesauthor[name={David Bauch}, color=blue]{db}
\definechangesauthor[name={Dustin Siebert}, color=red]{ds}
\definechangesauthor[name={Jens Förstner}, color=red]{jf}
\definechangesauthor[name={Klaus D. Jöns}, color=red]{kj}
\definechangesauthor[name={Stefan Schumacher}, color=red]{ss}

\usepackage{float}
\newcommand{\mueV}{\upmu\text{eV}}
\newcommand{\meV}{\text{meV}}
\renewcommand{\i}{\text{i}}
\renewcommand{\d}{\text{d}}

\newcommand{\XB}{\text{XX}}
\newcommand{\XS}{\text{X}}
\newcommand{\XH}{\text{H}}
\newcommand{\XV}{\text{V}}
\newcommand{\XG}{\text{G}}

\newcommand{\indist}{\mathcal{I}}
\newcommand{\conc}{\mathcal{C}}
\newcommand{\purcell}{\mathcal{F}_P}
\newcommand{\Q}{\mathcal{Q}}

\begin{document}

\preprint{APS/123-QED}

\title{On-demand indistinguishable and entangled photons \\using tailored cavity designs}

\author{David Bauch}
\affiliation{Department of Physics and Center for Optoelectronics and Photonics Paderborn (CeOPP), Paderborn University, Warburger Strasse 100, 33098 Paderborn, Germany}

\author{Dustin Siebert}
\affiliation{Electrical Engineering Department and Center for Optoelectronics and Photonics Paderborn (CeOPP), Paderborn University, Warburger Str. 100, 33098 Paderborn, Germany}

\author{Klaus D. J\"ons}
\affiliation{Department of Physics and Center for Optoelectronics and Photonics Paderborn (CeOPP), Paderborn University, Warburger Strasse 100, 33098 Paderborn, Germany}
\affiliation{Institute for Photonic Quantum Systems (PhoQS), Paderborn University, 33098 Paderborn, Germany}

\author{Jens F\"orstner}
\affiliation{Electrical Engineering Department and Center for Optoelectronics and Photonics Paderborn (CeOPP), Paderborn University, Warburger Str. 100, 33098 Paderborn, Germany}

\author{Stefan Schumacher}
\affiliation{Department of Physics and Center for Optoelectronics and Photonics Paderborn (CeOPP), Paderborn University, Warburger Strasse 100, 33098 Paderborn, Germany}
\affiliation{Institute for Photonic Quantum Systems (PhoQS), Paderborn University, 33098 Paderborn, Germany}
\affiliation{Wyant College of Optical Sciences, University of Arizona, Tucson, Arizona 85721, USA}

\date{\today}
             
\begin{abstract}
The biexciton-exciton emission cascade commonly used in quantum-dot systems to generate polarization entanglement yields photons with intrinsically limited indistinguishability. 
In the present work we focus on the generation of pairs of photons with high degrees of polarization entanglement and simultaneously high indistinguishability. We achieve this goal by selectively reducing the biexciton lifetime with an optical resonator. We demonstrate that a suitably tailored circular Bragg reflector fulfills the requirements of sufficient selective Purcell enhancement of biexciton emission paired with spectrally broad photon extraction and two-fold degenerate optical modes. Our in-depth theoretical study combines (i) the optimization of realistic photonic structures solving Maxwell's equations from which model parameters are extracted as input for (ii) microscopic simulations of quantum-dot cavity excitation dynamics with full access to photon properties. We report non-trivial dependencies on system parameters and use the predictive power of our combined theoretical approach to determine the optimal range of Purcell enhancement that maximizes indistinguishability and entanglement to near unity values, here specifically for the telecom C-band at $1550\,\mathrm{nm}$.
\end{abstract}

\maketitle

\section{\label{sec:introduction}Introduction}

Optical information carriers are used for data communication applications that demand high information bandwidth, immunity to electromagnetic interference, and long distance transmission. Emitters generating light for these purposes need to operate at a light frequency that minimizes the optical losses in the photon-carrying fiber, for example the telecom-C-band at around $\SI{1550}{nm}$.
Ideal candidates to generate photons at a desired wavelength are quantum dots (QDs) embedded in semiconductor microcavities \cite{Ding2016OnDemandSinglePhotons,schweickert2018demand}. These structures enable the generation of polarization controlled \cite{heinze2015quantum,jonas2022nonlinear}, highly indistinguishable, single photons \cite{Ding2016OnDemandSinglePhotons} and entangled photon pairs \cite{Chen2018EntangledPhotonsEfficientExtractionAntenna,Huber2008StrainEntangledOnDemand,huber2017highlySPTPSourceIndistConc,schumacher2012cavity,zeuner2021demand}. Additionally, the ideal QD-cavity system generates only a single set of photons per excitation cycle \cite{fischer2016dynamical,Hanschke2018SinglePhoton}, rendering them excellent quantum emitters for use in optical quantum information processing networks.


In the present paper we theoretically analyze the generation of pairs of photons with high degree of polarization entanglement and at the same time high values of indistinguishability. While in the biexciton-exciton cascaded emission of polarization entangled photons from semiconductor quantum dots the photon indistinguishability is intrinsically limited \cite{scholl2020crux}, here we systematically explore how this restriction can be alleviated by tailoring the photonic environment of the quantum dot emitter. Using a suitable photonic resonator structure, the biexciton to exciton transition can be selectively Purcell enhanced, reducing the effective biexciton lifetime, which reduces the temporal uncertainty in the correlated emission of the two photons. This in turn leads to increased photon indistinguishability \cite{scholl2020crux, PhysRevLett.128.093603}. Among the different photonic structures that can be used to optimize and enhance the photon emission are circular Bragg gratings (CBG) \cite{kolatschek2021bright, birowosuto2012fast, moczala2020strain, ji2021design}. These structures offer high light-matter coupling strengths, usually at relatively low $\Q$-values, combined with high photon collection efficiency over a broad spectral range \cite{PhysRevLett.122.113602,kolatschek2021bright}. Due to the rotational symmetry of the cavity, optical modes can exist in degenerate pairs with different polarizations allowing the generation of polarization entangled photons. While the general approach was initially implemented experimentally for a wavelength of $880\,\mathrm{nm}$ \cite{liu2019solid,PhysRevLett.122.113602}, an in-depth analysis of quantum-dot cavity photon sources where both the polarization entanglement and indistinguishability of the individual photons is systematically optimized has not been presented. 



Here we give a comprehensive theoretical investigation starting with the analysis of the cavity field of a realistic circular Bragg resonator by solving Maxwell's equations, using an optimization approach for 2D- and 3D-simulations of the physical structure. From that we extract parameters as input for density matrix theoretical simulations of the quantum optical excitation and emission dynamics of the quantum-dot cavity system. These calculations let us evaluate the emission characteristics of the structure, including photon indistinguishability and entanglement values. Based on this combined theoretical approach of Maxwell simulations and quantum optical QD-cavity simulations, we are then able to optimize our results towards experimentally implementable scenarios. We focus on the telecom C-band at $1550\,\mathrm{nm}$ and achieve indistinguishability and polarization entanglement values significantly above their reference values without selective Purcell enhancement, both approaching unity for optimized parameters.

\section{\label{sec:model}Emitter Structure}
We start by describing the emitter structure and its numerical treatment in further detail.
The temporal properties of the QD-cavity system and its corresponding emission dynamics is calculated using the von-Neumann equation \cite{bauch2021ultrafast}. Further detail of the theory is provided in \cref{sec:theory}.
The four level (4LS), diamond shaped system of the QD-biexciton is characterized completely by the exciton energy $E_\XS \approx \SI{0.8}{eV}$, the finestructure splitting energy $E_\text{FSP}=\SI{2}{\mueV}$ which separates the two different, linearly polarized excitons (H,V) and the biexciton binding energy, which typically extends up to several $\SI{}{meV}$ \cite{bauch2021ultrafast} and is set to a large but achievable $E_\text{Bind}=\SI{5}{\meV}$ for our simulations. The electronic states couple to a single-mode optical resonator (cavity) for each polarization mode, respectively. If not stated otherwise, the cavity energy $E_\text{cavity}$ is set equal to the biexciton-exciton transition energy such that $E_\text{cavity} = E_\XB-E_\XS \equiv \SI{1550}{nm}$ \cite{bauch2021ultrafast}.
\begin{figure}[t]
\includegraphics[width=\columnwidth]{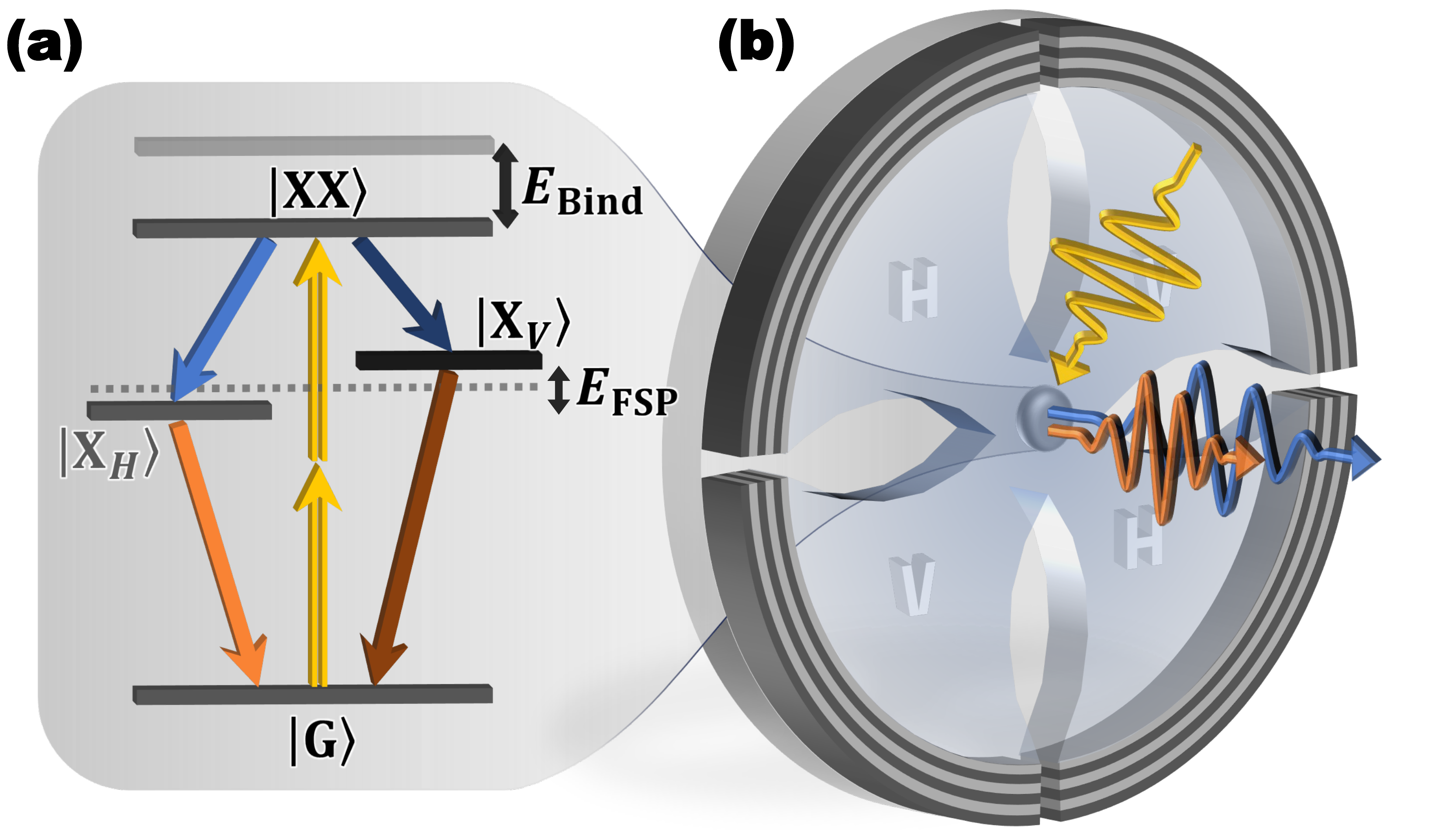}
\caption{\label{fig:physics}\textbf{(a)} Sketch of the QD-cavity structure with electronic configuration of the exciton-biexciton system as an inset, including the ground state $\ket{\XG}$, the two distinct exciton states $\ket{\XS_\XH}$, $\ket{\XS_\XV}$ and the biexciton $\ket{\XB}$. The excitons are energetically separated by the finestructure splitting $E_\text{FSP} = \left|E_{\XS,\XH}-E_{\XS,\XV}\right|$. The biexciton is redshifted by the binding energy $E_\text{Bind} = 2E_\XS - E_\XB$. \textbf{(b)} Artistic sketch of a CBG cavity with two distinct optical modes H,V; a realistic structure is shown in \cref{fig:maxwell_cavity_mode} below. The excitation pulse (yellow) generates maximum QD population. The QD emission from the biexciton-exciton transition (blue) and from the exciton-ground transition (orange) are highly indistinguishable, and the photons from both emission paths are polarization entangled.}
\end{figure}
\begin{figure}[t]
\includegraphics[width=\columnwidth]{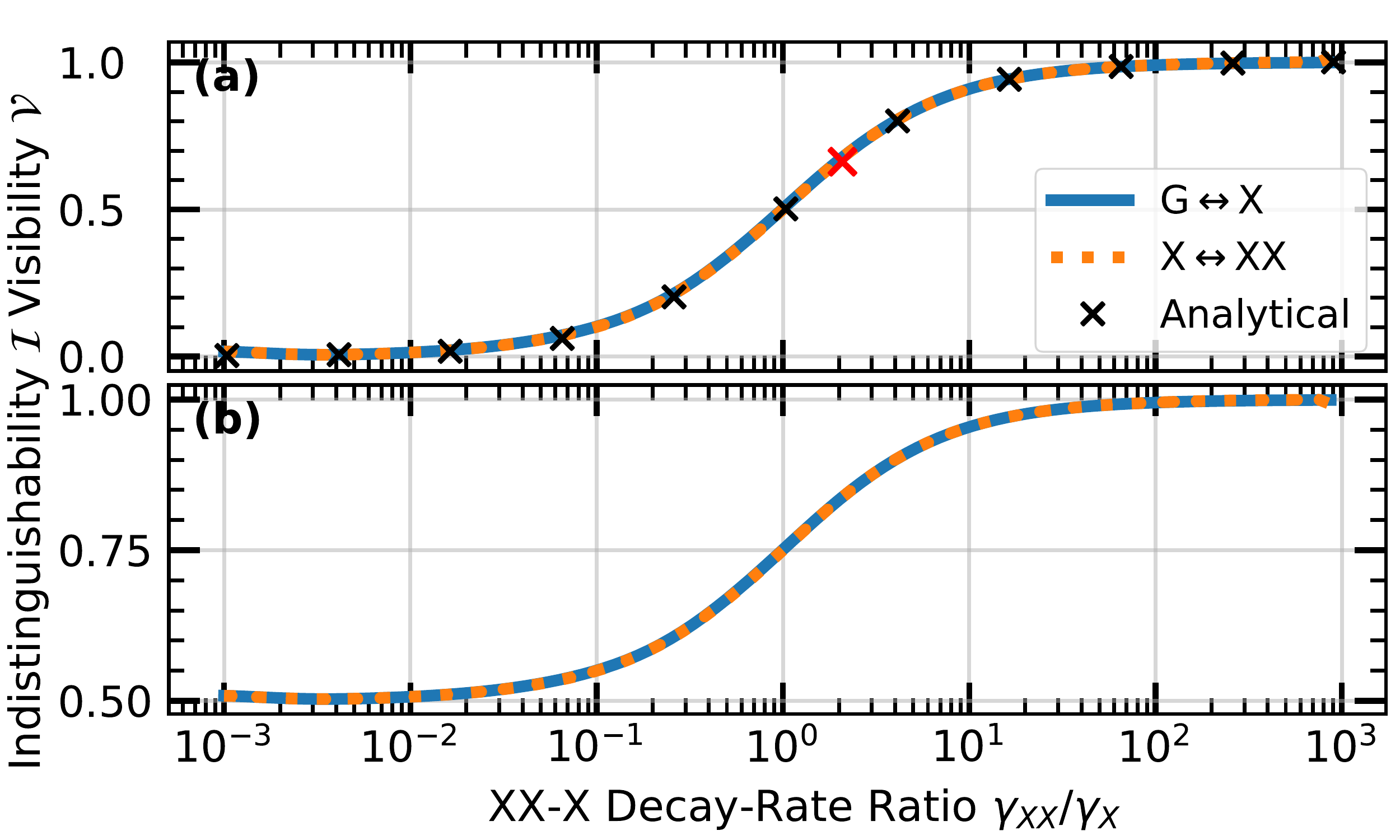}
\caption{\label{fig:poc_3ls}Visibility $\mathcal{V}$ \textbf{(a)} and indistinguishability $\indist$ \textbf{(b)} for a swept biexciton-exciton decay rate ratio $\gamma_\XB/\gamma_\XS$. The photons from the exciton-ground transition (blue solid lines) behave equally to the photons from the biexciton-exciton transition (orange dashed lines). The analytically estimated  visibility $\tilde{\mathcal{V}} = \left[1+\gamma_\XS/\gamma_\XB\right]^{-1}$ (black crosses; red cross marks $\gamma_\XB/\gamma_\XS=2$ for a free 3LS with no cavity) agrees well with the numerical simulation \cite{scholl2020crux}, indicating a large biexciton-exciton decay rate ratio is crucial for the generation of highly indistinguishable photons from the biexciton cascade.}
\end{figure}
It was previously shown in a simplified model, that the visibility of the emitted photons is increased by decreasing the biexciton lifetime. Respectively, the visibility is decreased when the lifetime ratio increases. \cref{fig:poc_3ls} compares the visibility curve as in \cite{scholl2020crux} with the indistinguishability from \cref{eqn:homindist}. The latter is a figure of merit much more inclusive to non-zero second order correlations $G^{(2)}$, which is intrinsically necessary in a two-photon emission. 
For a simple two level system (2LS) or three level system (3LS), no second order correlations are possible due to the limited Hilbert space. Thus, for the double time integrated correlation function $\Bar{\Bar{G}}^{(2)}\approx 0$ (\cref{eqn:double_time_integrated_g2}), the visibility (\cref{eqn:visibility}) becomes a sufficient figure of merit for the single photon indistinguishability. However, in this work, second order correlations contribute to a significant degree with $\Bar{\Bar{G}}^{(2)}\neq 0$. 
Thus, to include the relevant multi-photon processes in the emission cascade, the indistinguishability $\indist$, as defined in the appendix in \cref{eqn:homindist}, will be utilized. 
For the isolated biexciton with no cavity, according to \cref{eqn:visibility,eqn:homindist}, the visibility for the spontaneously emitted, entangled photons is limited to $\mathcal{V}_\text{Ref} = 0.66$ \cite{scholl2020crux}, which corresponds to the indistinguishability $\indist_\text{Ref}=0.82$. While higher values are achievable without a cavity only by e.g. choosing different semiconductor materials \cite{huber2017highlySPTPSourceIndistConc} or by stimulated emission using optical driving of the QD \cite{PhysRevLett.128.093603}, these structures are typically strongly limited in the repetition rate of the emitter due to the long lifetimes of the QD states. The tuning of the biexciton lifetime has been demonstrated experimentally by using a control laser to stimulate the biexciton-exciton emission \cite{sbresny2022stimulated}, greatly reducing the biexciton lifetime and increasing the indistinguishability of the emitted photon.

While the main focus of the present paper is on tailoring the emission process from the QD-biexciton, the biexciton may be excited using several approaches, each of which offers different advantages and constraints. 
The specific type of excitation potentially also reduces the quality of the emitted photons. As an example, the direct two-photon excitation sets a fundamental limit to the achievable entanglement \cite{seidelmann2022two}.
In order to isolate the effect of the Purcell-enhanced biexciton-exciton transition, in the first part of the analysis we will  assume a perfectly prepared biexciton at $t = 0$. 
Later we will come back to the aspect of biexciton initialization and demonstrate that it does not significantly influence or even undermine our strategy for the polarization entangled photon source optimization. 

\begin{figure}[t]
\includegraphics[width=\columnwidth]{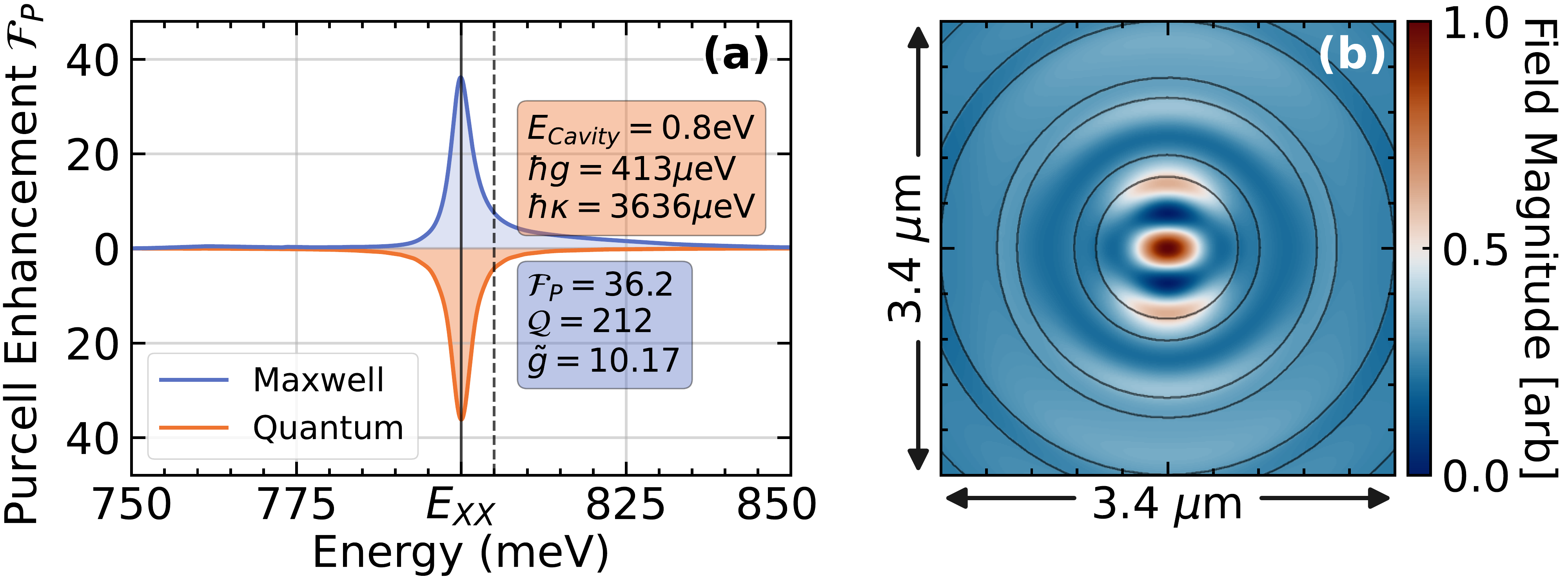}
\caption{\label{fig:maxwell_optimized_cavities} \textbf{(a)} Cavity emission spectra for the Maxwell simulations (blue line) and the QD simulation (orange line). Low-$\Q$ CBG cavity optimized to around $\SI{0.8}{eV}$. For the inset parameters, the emission spectrum calculated using the Maxwell simulations shows an almost isolated emission at the desired frequency with only small side modes. Though much weaker, the exciton (black dashed line) is also enhanced by the cavity. The coupling $\tilde{g}$ translates to the cavity coupling by $\hbar g = \hbar \tilde{g}\sqrt{\gamma_\text{Rad}}/\sqrt{\text{ps}}$. \textbf{(b)} The normalized fundamental field mode corresponding the the spectrum in (a). The intensity is strongest at the center of the reflector, requiring precise positioning of the QD to maximize the light-matter coupling and thus the cavity coupling rate $\hbar g$.}
\end{figure}

\section{\label{sec:cavity}Maxwell Simulations}
\begin{figure}[t]
\includegraphics[width=\columnwidth]{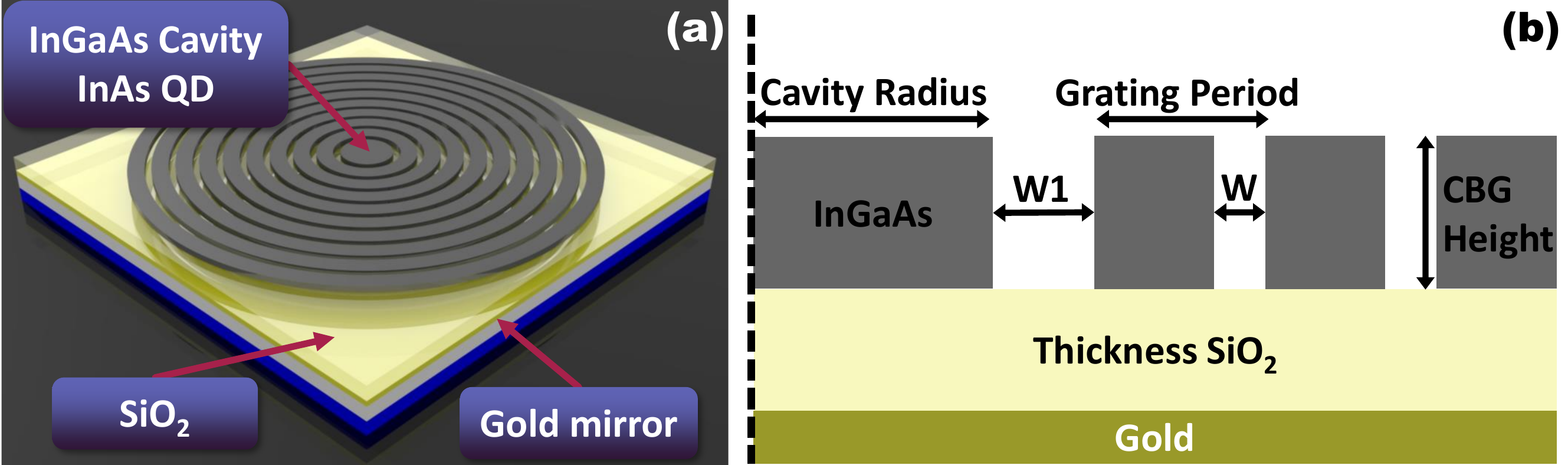}
\caption{\label{fig:maxwell_cavity_mode}\textbf{(a)} Rendered image of the simulated circular Bragg gratings. The InGaAs resonator structure is applied to a SiO\textsubscript{2} substrate. \textbf{(b)} Numerical treatment of the cavity. A sideview shows the symmetric cavity structure around its center (black dashed line). The cavity radius, grating period, first and remaining trench widths (W1 and W), the grating height and the thickness of the SiO\textsubscript{2} substrate can be tuned by the optimizer towards a target wavelength, coupling, $\Q$-factor and Purcell enhancement.}
\end{figure}

In the first step, we turn our attention to the optimization of the photonic structure. This optimization is, however, already aware of the results of our simulations of the QD-cavity emission dynamics as discussed in the next chapter. With the knowledge acquired there, we strive to obtain a cavity design with high coupling strength, moderate Purcell enhancement and low $\Q$-factor. In this regard, we introduce a new powerful optimization scheme for indium gallium arsenide (InGaAs) based CBG resonators in this work. The CBG resonator consists of a central disk defining the cavity for an InAs-QD surrounded by a radial Bragg reflector, which is underlaid with an SiO\textsubscript{2} layer and a gold mirror as shown in \cref{fig:maxwell_cavity_mode},a. Our device design was inspired by other recent publications, which investigated a similar structure both theoretically and experimentally \cite{bremer2022numerical, PhysRevLett.122.113602, moczala2020strain, Rickert:19}. Here, we choose an emission wavelength of \SI{1550}{\nano \meter} exhibiting minimal losses for today's fiber infrastructure, but we pursue the idea to achieve this target wavelength based on the already well investigated InGaAs material system. It has been already shown, that those InGaAs-CBG resonators, which are nothing else than circular photonic crystal cavities, can be accurately tuned to $\lambda_0 = \SI{1550}{\nano \meter}$ while being optimized for a maximum Purcell enhancement \cite{bremer2022numerical}. But as mentioned before, here we need another goal function for the QD-cavity emission dynamics. Thus, for our approach tailoring the CBG resonator parameters, we developed a new cascaded optimization scheme solving Maxwell equations. First, we solve a batch of low-numerical-cost 2D-simulations in the Finite Difference Time Domain (FDTD) Maxwell solver Meep \cite{johnson2010} utilizing the rotational symmetry of the structure. The results of this batch are then used for global Bayesian optimization via local penalization \cite{gonzalez2016, gpyopt2016}. As Bayesian optimization appears to one of the most powerful state-of-the-art methods for costly objectives of nanophotonic devices \cite{schneider2019}, the batch based approach makes it even more powerful, since solving a batch of simulations can be realized perfectly parallel [\cref{subsec:OptDetails}]. After we have obtained a partially converged result, we use it as initial guess for further local optimization polishing the former global optimum. We do this initially in a sequential 2D-FDTD Nelder-Mead optimization and finally port the result to a 3D-Finite Integration Technique (FIT) local optimization with the help from GPU accelerators \cite{CSTMWS2023}. On the one hand, this cascaded scheme prevents us from missing the actual optimum, since Bayesian optimization based on expected improvement acquisition functions mostly ensure finding the region around the global optimum instead of converging against one final, accurate solution \cite{Frazier2018}. In our studies, it was not possible to achieve the same convergence for the global Bayesian optimizer compared to the local ones based on the Nelder-Mead simplex method with an acceptable number of iterations. On the other hand, this scheme avoids significant deviations between 2D- and 3D-simulations in the structure's dimensions for a later fabrication process [\cref{subsec:maxFP}].



As optimization parameters depicted in \cref{fig:maxwell_cavity_mode},b we chose the CBG height, the radius of the central disk, the grating period, the grating's trench width and the SiO\textsubscript{2} layer thickness, respectively. Additionally, we added the width of the first trench as another degree of freedom, since the circular Bragg condition is only converging against the rectangular one with increasing ring count from the inside to the outside of the CBG. Therefore, the first trench width is expected to be larger than the remaining ones, as the field solutions of Maxwell equations for the cavity modes consist of a superposition of Bessel functions \cite{scheuer2003}.

With this optimization setup, defining objectives for maximum Purcell enhancement leads to unsuitable results for the creation of indistinguishable and entangled photons due to the fact the CBG cavity is showing Purcell factors above 360 [\cref{subsec:maxFP}]. Much more promising instead is an objective rewarding high coupling factors accurately tuned to the target wavelength with low $\Q$-factors and moderate Purcell enhancement on the other hand. The final optimization result for the cavity mode is shown in \cref{fig:maxwell_optimized_cavities},a. From \cref{eqn:coupling_em} we can connect the Maxwell simulations to our quantum simulations by choosing a fixed radiative decay rate and derive a convenient optimization criterion for the cavity exhibiting as high as possible indistinguishability and concurrence, as discussed in the following chapter.

\section{\label{sec:results}Quantum Simulation}
\begin{figure}[t]
\includegraphics[width=\columnwidth]{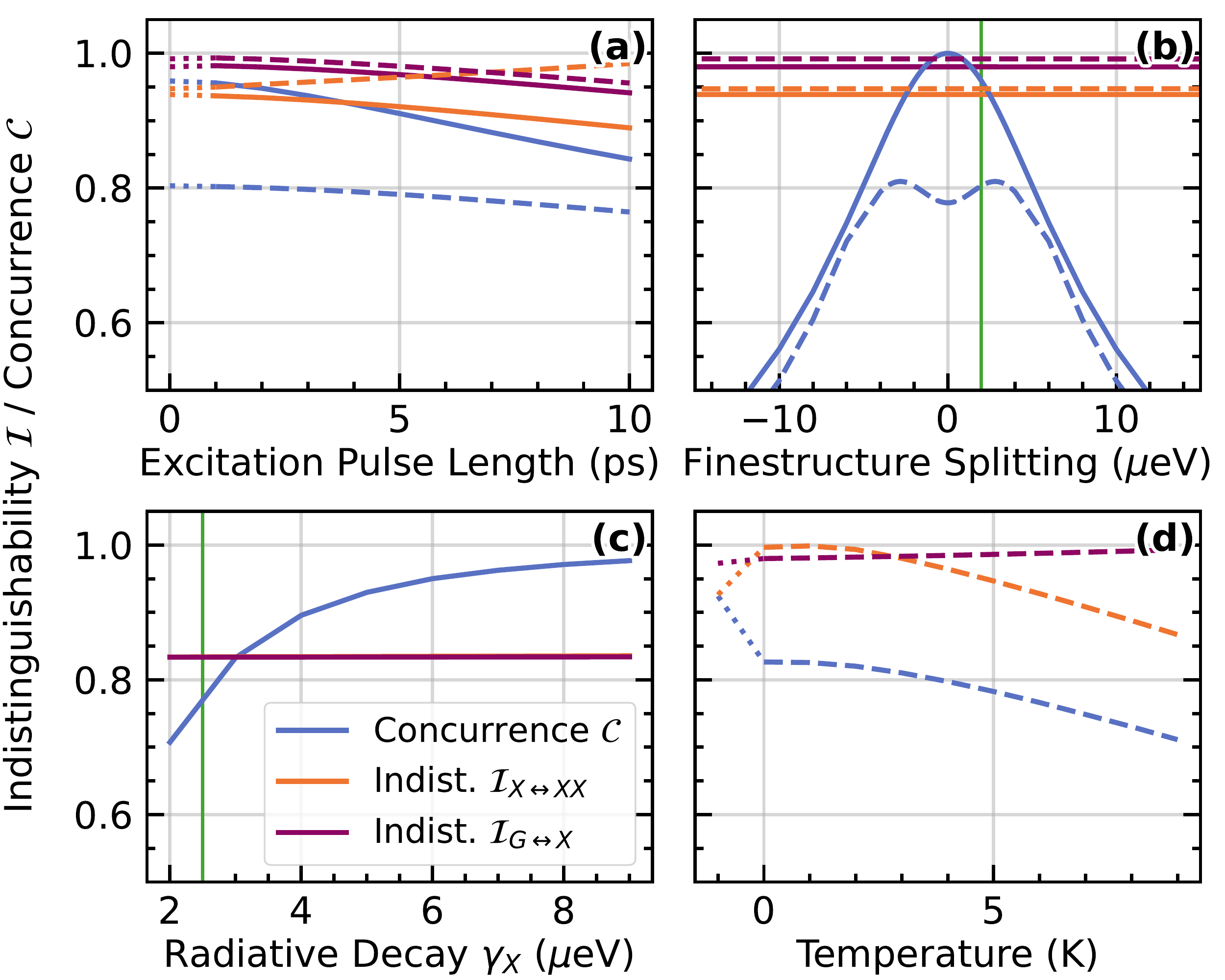}
\caption{\label{fig:purcell_excitation_and_reference}Quantum Properties for $\hbar g = \SI{200}{\mueV}$, $\hbar \kappa = \SI{3}{\meV}$ and $\gamma_\text{Rad} = \SI{2.5}{\mueV}$. The rate of cavity losses is equivalent to a $\Q$-factor $= 265$. The Purcell enhancement for these parameters results in $\purcell=11.5$. The cavity is again set to exactly the biexciton-exciton transition energy. For the solid lines, no electron phonon coupling is included. For the dashed lines, the temperature is fixed to $T = \SI{4.2}{K}$. \textbf{(a)} Quantum properties for a pulse-initialized biexciton. The dotted lines mark the reference values when simulating the system for a fully initialized biexciton $\rho(t=0)=\ketbra{\XB}{\XB}$. Otherwise, the initial state is set to $\rho(t=0)=\ketbra{\XG}{\XG}$. The pulse area is set to $\Omega = \SI{1}{\pi}$. The pulse width $\tau$ is varied in a range suitable for the experiment. \textbf{(b)} Quantum properties for different finestructure splitting energies. At $E_\text{FSP}=0$ the precession of the exciton states is zero, resulting in maximum concurrence, which is only reduced by the electron-phonon induced dephasing. \textbf{(c)} Quantum properties for an isolated biexciton decay for different rates of radiative decay. The indistinguishabilities remain constant for all rates of decay, while the concurrence exhibits a strong dependency where it is greatly reduced for small rates of radiative decay. No cavity is used here, providing a baseline for the cavity simulations. The green line marks the reference values for the cavity simulations in this work, where $\gamma_\text{Rad} = \SI{2.5}{\mueV}$. Both indistinguishabilities overlap. \textbf{(d)} Temperature dependency for a pulse width fixed to $\tau = \SI{4}{ps}$ where the temperature is varied between $\SI{0}{K}$ to $\SI{10}{K}$. The dotted lines mark the reference values when simulating the system without electron-phonon coupling.}
\end{figure}

For the emission scheme proposed in this work, the tuning of the cavity to approximately the biexciton-exciton resonance is required. We set the exciton energies such that $E_\XB-E_\XS = \SI{0.8}{eV}$, which corresponds to the $\SI{1550}{nm}$ telecom C-band. The Maxwell simulations suggest this is in theory possible using the proposed CBG cavity.
We numerically simulate the quantum mechanical dynamics using a two-time von-Neumann approach including non-Markovian losses with a Lindblad method. For details see appendix \cref{sec:theory}.
The reference values for the indistinguishability and concurrence are calculated for the emission from an isolated quantum dot with no cavity. Their numerical values remain at $\indist_\text{Ref} \approx 0.82$ for both of the single photons at any radiative decay rate and $\conc_\text{Ref} \approx 0.72$ for the two-photon entanglement at $\gamma_\text{Rad}=\SI{2.5}{\mueV}$. Unlike the indistinguishability, the concurrence is strongly dependent on the rate of radiative decay due to the precession of the exciton states \cite{ward2014coherent} and will increase with larger $\gamma_\text{Rad}$ (compare \cref{fig:purcell_excitation_and_reference},c) and decrease with rising finestructure splitting (compare \cref{fig:purcell_excitation_and_reference},b).
All parameter sets are simulated for times $t\in[0,\SI{1.5}{ns}]$. Because the biexciton is supposed to decay much faster than the exciton, for the $G^{(i)}(t,\tau)$ correlation functions (see \cref{eqn:g1,eqn:g2}), the $t-\tau$ grid is configured such that the first \SI{200}{ps} are evaluated using $\Delta t = \SI{100}{fs}$, while the remaining $\SI{1.3}{ns}$ are evaluated using $\Delta t = \SI{500}{fs}$, which is sufficient to obtain numerically converged results.

\begin{figure}[t]
\includegraphics[width=\columnwidth]{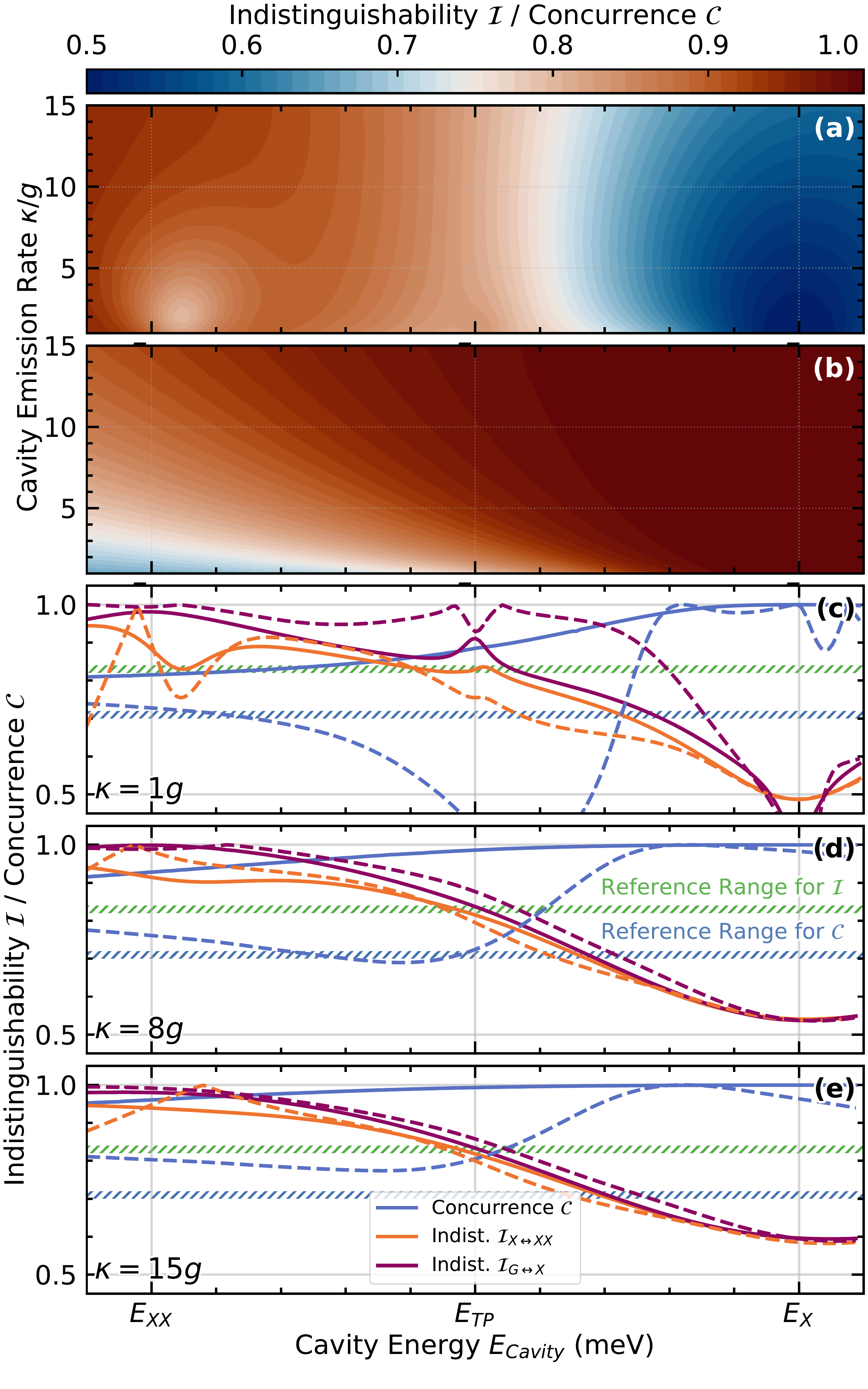}
\caption{\label{fig:poc_biexciton}Indistinguishability $\indist$ \textbf{(a)} and Concurrence $\conc$ \textbf{(b)} for varying cavity energy $E_c = \hbar\omega_c$ and cavity emission rate ratio $\kappa / g$. The QD-cavity coupling is set fixed to $\hbar g = \SI{200}{\mueV}$. The biexciton binding energy is arbitrarily set to $E_\text{Bind} = \SI{5}{\meV}$. The simulation starts with a fully initialized biexciton state without any cavity photons $\rho_0 = \ketbra{\XB}$, effectively simulating a perfect excitation cycle followed by the ideal emission for a given set of parameters. The radiative decay rate is set to $\hbar\gamma_\text{Rad} = \SI{2.5}{\mueV}$, corresponding to an exciton lifetime of $\tau_\XS = 1/\gamma_\text{Rad} \approx \SI{260}{ps}$. Cross-sections at $\kappa=g$, $\kappa=8g$ and $\kappa=15g$ \textbf{(c-e)} include results without further losses (solid lines) and with active electron-phonon coupling at $T = \SI{4.2}{K}$ (dashed lines). The reference value for the indistinguishabilities (hatched green area) is $\indist_\text{Ref}\approx 0.82$ and for the concurrence (hatched blue area) $\conc_\text{Ref}\approx 0.72$.}
\end{figure}
In \cref{fig:poc_biexciton}, we sweep the cavity frequency and the cavity losses. The latter directly translates into the cavity-$\Q$-factor \cref{eqn:purcell_factor_qd} for $E_c = \SI{0.8}{eV}$.
For the cavity at the exciton-ground transition, the photon emission results in highly entangled photons. This is the direct result of the very short exciton lifetime while the biexciton lifetime remains high. When the biexciton decays, so does the exciton immediately too, leaving both photons strongly entangled. Because of the large lifetime differences, the indistinguishabilities remain low.
For the cavity at the two-photon resonance, the emission process resembles the reference emission with slightly shorter lifetimes of both the biexciton and the exciton. As has been demonstrated previously \cite{bauch2021ultrafast,heinze2017polarization}, the entanglement may remain very high for this setup. The indistinguishabilities are reduced to their corresponding reference values, especially for low-$\Q$ cavities.
For the cavity at the biexciton-exciton transition, the emission changes drastically. For all investigated $\Q$-values, the indistinguishabilities of the biexciton-exciton photon and the exciton-ground photon is greatly increased. This is due to the decreased biexciton lifetime, which allows the biexciton to decay on timescales where its ground state, namely the excitons, remains stable.
The biexciton-exciton indistinguishability exhibits a slightly more complex behaviour than its exciton-ground counterpart, where $\indist_\XB$ is decreased directly before the total resonance condition is reached. Exceeding the resonance threshold and moving to slightly lower energies, the indistinguishability is maximized. 

\begin{figure}[t]
\includegraphics[width=\columnwidth]{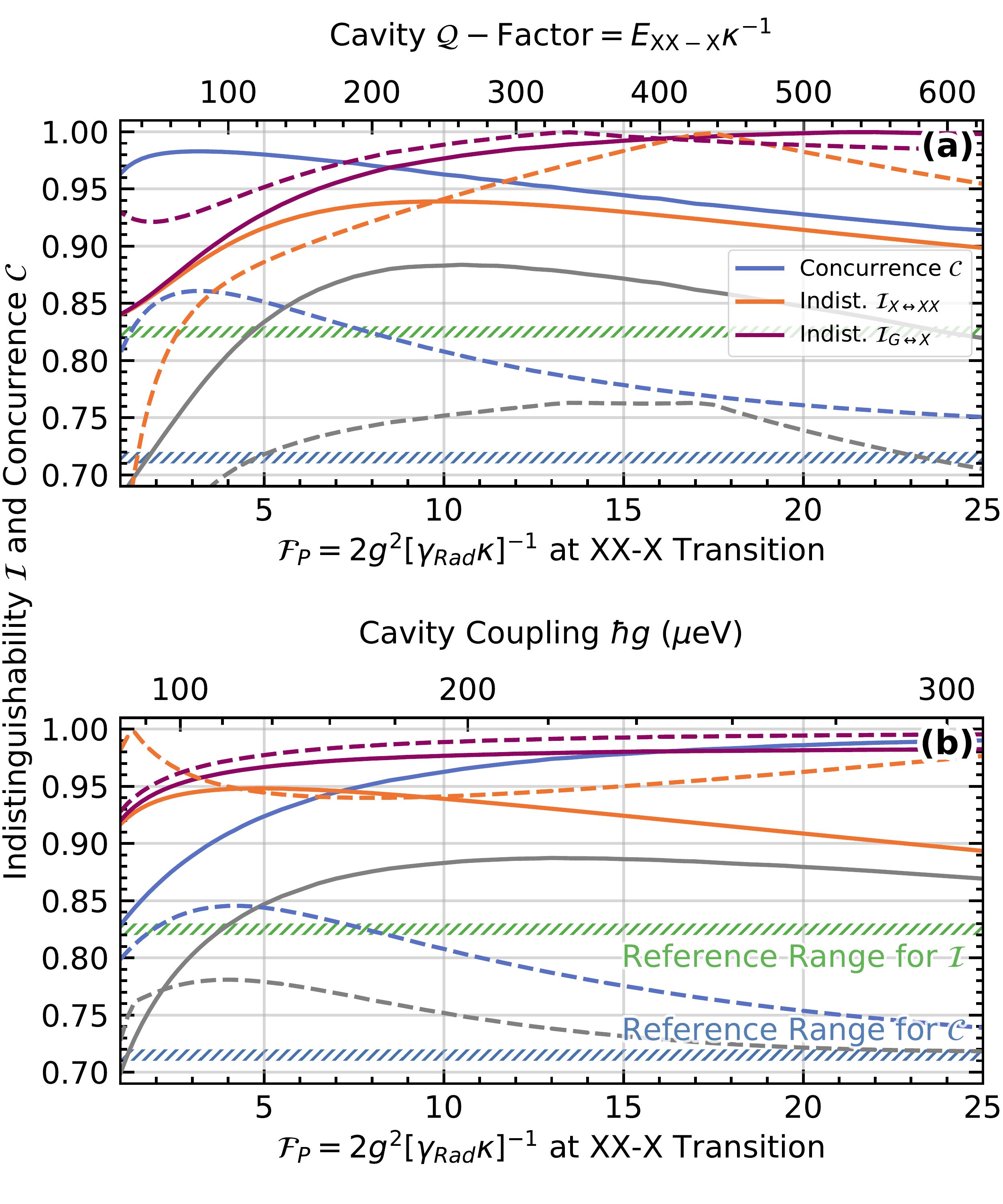}
\caption{\label{fig:purcell_scan}Photon properties for varying Purcell factor, which can be tuned adjusting either the cavity $\Q$-factor \textbf{(a)} (fixed $\hbar g =  \SI{200}{\mueV}$), or the light-matter coupling \textbf{(b)} (fixed $\hbar\kappa = \SI{3}{\meV}$). The radiative decay rate is set fixed to $\gamma_\text{Rad} = \SI{2.5}{\mueV}$, the cavity is set to exactly the biexciton-exciton transition energy, such that $E_\text{Cavity} = E_\XB - E_\XS$. The simulation is initialized with a fully excited biexciton. The concurrence (blue) and both the exciton-ground indistinguishability (violet) as well as the biexciton-exciton indistinguishability (orange) are multiplied into a figure of merit where all properties are maximized (grey). Results with (dashed lines) and without (solid lines) electron phonon coupling are presented. The reference values for the indistinguishabilities (hatched green area) group around $0.82$. The reference value for the concurrence (hatched blue area) remains at around $0.72$ for the set rate of radiative decay.}
\end{figure}
In \cref{fig:purcell_scan} we sweep the Purcell factor from \cref{eqn:purcell_factor_qd} by tuning either the cavity loss rate $\kappa$ or the coupling constant $g$ while setting the respective other parameter fixed. The resulting $\purcell$-factor ranges between $1$ and $25$, which is realistic for the cavities described in \cref{sec:cavity}.
For the modulation of $\purcell$ using the cavity losses in \cref{fig:purcell_scan},a, optimal results are achieved for a Purcell factor of $\purcell\approx 5$ to $25$. 
The single photon indistinguishability of the biexciton-exciton photon $\indist_\XB$ is maximized at low-$\Q$ cavities for a Purcell factor ranging from $\purcell \approx 5$ to $10$. High indistinguishability of the exciton-ground photon $\indist_\XS$ is achieved for high-$\Q$ cavities. Fortunately, even for low-$\Q$ cavities where $\indist_\XB$ is maximized, $\indist_\XS$ is still surpassing its corresponding reference value of $0.82$. Contrary, the concurrence $\conc$ is maximized for $\purcell\rightarrow 2.5$, while also exceeding its reference value for low-$\Q$ cavities. The global optimum where all of these quantities are maximized equally is achieved for $\purcell\approx 10$ (\cref{fig:purcell_scan},a,grey line). For higher-$\Q$ cavities, all properties investigated tend to plateau.
These results hold even with electron phonon coupling enabled at $T = \SI{4.2}{K}$. At low temperatures, the indistinguishabilities even increase slightly for $\purcell > 10$ due to the accelerated decay of the QD-population when compared to the phonon-free simulation, up to reaching near unity values at $\purcell \approx 17$. Thereafter, the phonon induced dephasing dominates, and the indistinguishabilities are reduced again. The entanglement mainly suffers from the phonon induced dephasing, where a general reduction as well as a stronger decrease with rising $\purcell$ occurs.
Similar results are achieved when varying $\purcell$ using the cavity coupling rate, which is depicted in $\hbar g$ \cref{fig:purcell_scan},b. Here, the biexciton single photon indistinguishability $\indist_\XB$ is maximized at $\purcell\approx 4$, while the global optimum is set around a slightly higher $\purcell \approx 5$ to $15$.
For larger cavity couplings, the entanglement as well as the exciton-ground photon indistinguishability are maximized towards unity, suggesting a cavity design with strong couplings $\hbar g\geq\SI{200}{\mueV}$ is optimal for the implementation of this type of emitter. According to our Maxwell simulations in \cref{sec:cavity}, these values are perfectly achievable using CBG cavities.
Overall, a reduction of the cavity coupling $\hbar g$ and losses $\hbar\kappa$ to a unified Purcell factor is not sufficient to maximize all of the indistinguishabilities and the photon entanglement simultaneously. While increases from their corresponding reference values are trivially possible for a large range of parameters, maximizing all three properties requires precise tuning of the couplings, losses, frequencies and ideally the temperature. Instead, aiming for double digit Purcell enhancements for the biexciton-exciton transition, large light-matter coupling strengths and low-$\Q$ cavities are desirable.

Because the excitation process can intrinsically lower the quantum properties of the emitted photons, notably the two-photon entanglement \cite{seidelmann2022two}, no excitation process was included in the simulations shown thus far. To determine whether the proposed emission scheme holds in a realistic environment where the biexciton preparation cannot be neglected, a set of parameters where the quantum properties exhibit high values are tested with the excitation process included. As briefly discussed already above, different optical excitation schemes to populate the biexciton are available, each coming with its own potential advantages and disadvantages. Direct, coherent excitation using a single $\pi$-pulse requires spectral separation as well as polarization filtering of the excitation laser and the QD emission \cite{ramsay2010review}. Off-resonant, phonon-assisted excitation at low temperatures allows for high excitation efficiencies, evading the laser-background problem \cite{glassl2013proposed,quilter2015phonon,ardelt2014dissipative}. Furthermore, using multiple laser pulses in a degenerated two-photon process \cite{ardelt2014dissipative,stufler2006two,jayakumar2013deterministic} or two-color dichromatic excitation strongly detuned from the QD transitions \cite{bracht2021swing,karli2022super} also allow for high excitation fidelities. For a Purcell-enhanced transition via a semiconductor cavity, the excitation fidelity of this process may be greatly decreased due to the re-excitation of the QD and premature emission of the photons. This problem may be circumvented using cavity-off-resonant excitation of the QD. Here, not only the excitation laser but also the cavity is strongly off-resonant to the QD transitions \cite{bauch2021ultrafast}. 

In order to keep the excitation process simple, here we use a Gaussian shaped pulse to directly generate biexciton population via the two-photon Rabi-flop absorption process. Varying the pulse length from $\tau\in[\SI{1}{ps},\SI{10}{ps}]$, the emission process yields high quality photons throughout the investigated range of pulse widths. The resulting quantum properties are displayed in \cref{fig:purcell_excitation_and_reference},a. Without electron-phonon coupling, the indistinguishabilities are reduced only slightly for larger pulse widths. The concurrence is reduced more severely while still remaining above the reference value, even for long pulses.
Furthermore, varying the temperature for a simulation including the excitation process, which is depicted in \cref{fig:purcell_excitation_and_reference},c, the quantum properties are stable for very low temperature $T \leq \SI{2}{K}$. Here, the phonon enhanced decay of the QD-states can even result in slight improvements for the quantum properties. For larger temperatures, the phonon induced dephasing becomes dominant again, significantly reducing the biexciton-exciton indistinguishability as well as the entanglement.

\begin{figure}[t]
\includegraphics[width=\columnwidth]{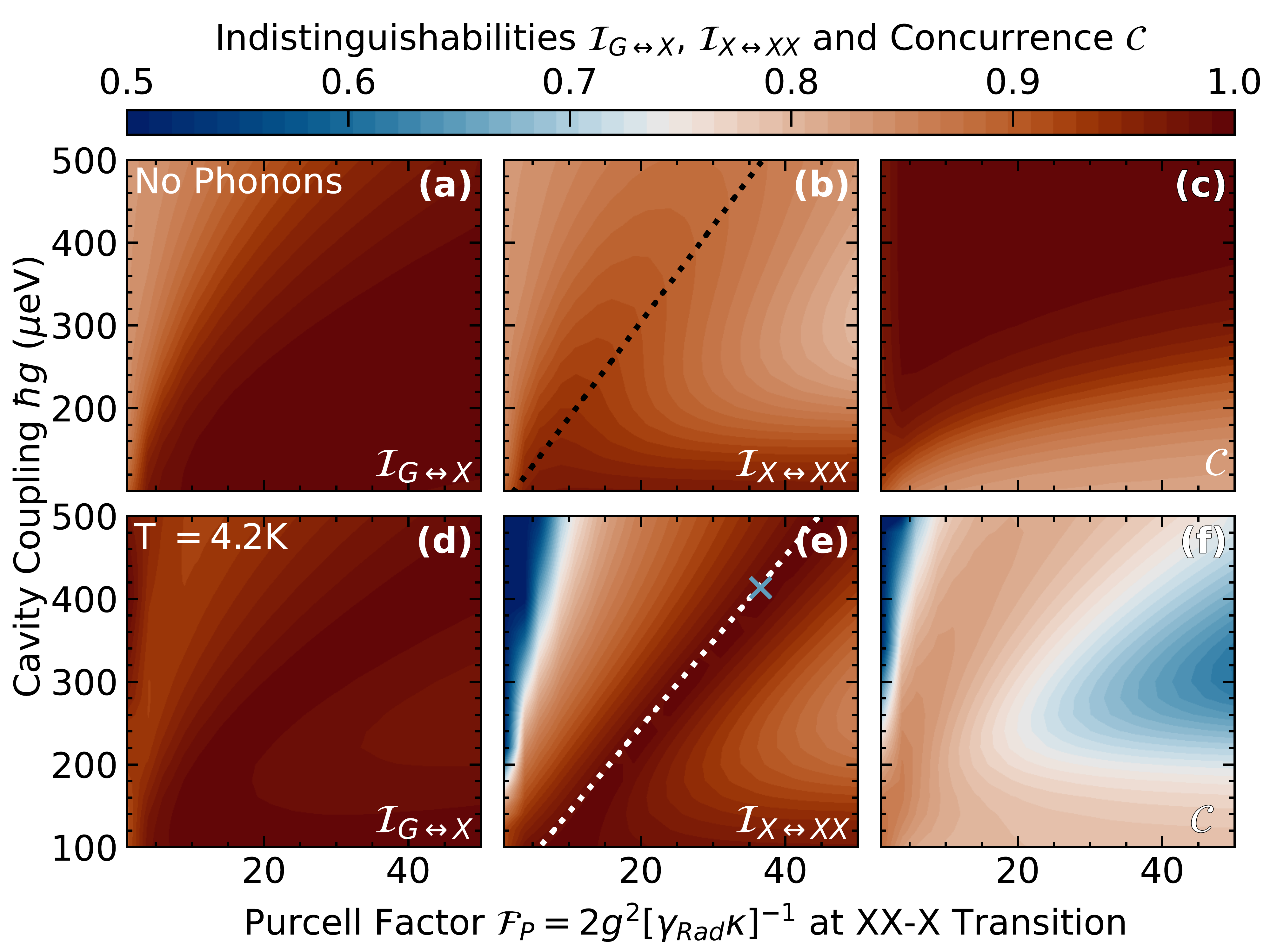}
\caption{\label{fig:g_kappa_sweepT4.2k}Indistinguishability of the exciton photon $\indist_\XS$ \textbf{(a,d)}, of the biexciton photon $\indist_\XB$ \textbf{(b,e)} and their concurrence $\conc$ \textbf{(c,f)} for a swept cavity coupling $\hbar g \in[\SI{100}{\mueV},\SI{500}{\mueV}]$ and Purcell factor $\purcell\in[1,50]$. The radiative decay rate is set to $\SI{2.5}{\mueV}$ and the temperature is set to $T=\SI{4.2}{\kelvin}$ (d-f, no electron phonon coupling in a-c). The cavity is set to exactly the biexciton-exciton transition. A linear fit for the local maxima of the 2D data (white dotted line) reveals the maxima of the product of all the quantum properties depicted to follow $\hbar g =\alpha \purcell + \beta$ with $\alpha \approx 10.3~(11.6)$ and $\beta \approx 38~(72)$ (with no electron phonon coupling, black dotted line). The blue cross in (e) corresponds to the simulated cavity in \cref{fig:maxwell_optimized_cavities},a.}
\end{figure}

Because a low $\Q$-factor and high light matter coupling constant $\hbar g$ appear to be required for generating strongly entangled and individually highly indistinguishable photons, a broad sweep of both parameters is evaluated in \cref{fig:g_kappa_sweepT4.2k}. Here it becomes clear that maximizing all properties at the same time is possible  over a large range of parameters. 
The indistinguishability of the exciton photon-ground $\indist_\XS$ is maximized for high coupling, high-$\Q$ cavities. 
The indistinguishability of the biexciton photon $\indist_\XB$ is maximized for high coupling, low-$\Q$ cavities, where a specific Purcell enhancement has to be achieved. 
The concurrence of these photons is also maximized for high coupling, low-$\Q$ cavities, leaving these cavities as the ideal choice when tolerating small losses in exciton indistinguishability.

These results are illustrated in \cref{fig:g_kappa_sweepT4.2k}. 
The threshold for the Purcell enhancement for which these properties are maximized appears to scale with the cavity coupling rate $\hbar g$. For instance, for the $\hbar g = \SI{200}{\mueV}$ cavity investigated in \cref{fig:purcell_scan} and \cref{fig:poc_biexciton}, a Purcell enhancement of $\purcell \in [5,15]$ maximized these values. For larger couplings, higher $\purcell$ are necessary, as can be seen in the linear dependency of $\indist_\XB$ in \cref{fig:g_kappa_sweepT4.2k}. Thus, larger couplings necessitate larger enhancement thresholds.

We may also fit a linear dependency of the light-matter coupling and the Purcell enhancement to $\hbar g \approx \alpha\purcell + \beta$, where for our parameters and for $T=\SI{4.2}{K}$, $\alpha\approx \SI{10.3}{\mueV}$ and $\beta\approx \SI{38}{\mueV}$ (compare \cref{fig:g_kappa_sweepT4.2k},b,e) 
for the QD-cavity system to yield the best possible photons. The scaling factor $\alpha$ depends on the strength of the electron-phonon coupling and therefore the temperature (see \cref{subsec:tempdep}). This information can then be used to refine the fitness function for the Maxwell optimizer described in \cref{sec:cavity}, resulting in precise tuning of the required cavity mode to maximize the photon quality.


\section{\label{sec:conclusion}Conclusion}
We have presented an in-depth theoretical and numerical analysis of a quantum emitter generating highly indistinguishable and simultaneously entangled photons.
While photon emitters at such frequencies have already been realized experimentally \cite{birowosuto2012fast}, the indistinguishability without Purcell enhancement of the biexciton-exciton transition is intrinsically limited by the biexciton-exciton lifetime ratio \cite{scholl2020crux}. This has been a limiting restriction for application in quantum information technologies, where indistinguishability and entanglement is key for the implementation of quantum protocols such as entanglement swapping.
We generate the entangled photon pairs, where the individual photons are themselves indistinguishable,  using a cavity with resonance frequency set around the biexciton-exciton transition.
The cavity ensures a Purcell enhancement of the biexciton-exciton transition, significantly reducing the biexciton lifetime while leaving the exciton lifetime constant.
A realistic structure design was achieved based on a combined approach of Maxwell simulations and density matrix simulations of the quantum-dot cavity system, confirming the viability of CBG cavities for our purpose.
We optimized the cavity structure to favour low $\Q$-factors and large coupling values with low Purcell enhancement. Notably, our results show that optimization towards large Purcell enhancements, as is customary for these structures \cite{Rickert:19,bremer2022numerical}, is not equally advantageous.
The reduction of the biexciton lifetime and thus the removal of the intrinsic limit of the indistinguishability then results in photons reaching indistinguishabilities with $\indist>95\%$, while leaving the photon entanglement at high values with $\conc>95\%$.
For the investigated set of cavity parameters, a Purcell enhancement of $\purcell\in[5,15]$ appears to be ideal, where the exact value depends on the details of the individual quantum dot, most importantly the rate of radiative decay and the temperature. 
For low temperatures with included electron phonon coupling, our simulations show further improvements in the indistinguishability of the biexciton photon. While still remaining a source of loss for the system due to the phonon induced dephasing, optimizing the parameters of the QD to obtain the best possible photon quality includes tuning of the temperature as well.
Similarly, an excitation method with maximum biexciton preparation fidelity is preferred, but not required for this emission scheme to yield high quality photons. 
However, the specifics of the excitation process depend strongly on the cavity, where utilizing advanced excitation methods like pulse shaping \cite{praschan2022pulse} or the SUPER scheme \cite{karli2022super,heinisch2023swing} may prove beneficial.
While our results are tailored towards the telecom wavelength at $\SI{1550}{nm}$, other wavelengths can be achieved with little further adjustment due to the broad cavity mode.
%

\begin{acknowledgments}
This work was supported by the Deutsche Forschungsgemeinschaft (DFG, German Research Foundation) through the transregional collaborative research center TRR142/3-2022 (231447078, projects B06 and C09), the German Federal Ministry of Education and Research (BMBF) via the project QR.X (No.16KlSQ012), the Photonic Quantum Computing initiative (PhoQC), and computing time provided by the Paderborn Center for Parallel Computing, PC$^2$. 
\end{acknowledgments}

\appendix

\section{\label{sec:theory}Theory}
We numerically evaluate the biexciton-cavity system using the von-Neumann equation 
\begin{align}
    \frac{\text{d}\rho}{\text{d}t} = \frac{\i}{\hbar}\left[\mathcal{H},\rho\right] + \sum \mathcal{L}_{\hat{O}}(\rho) \label{eqn:vonneumann}
\end{align}
with the biexciton-cavity Hamiltonian \cite{bauch2021ultrafast,PhysRevB.105.045302} in the interaction frame and with a rotating frame approximation
\begin{align}
    \mathcal{H} = &\sum_{i=H,V}g\left[ \ketbra{G}{X_i}\hat{b}_i^\dagger + \ketbra{X_i}{B}\hat{b}_i^\dagger \right] + \text{H.c.} \\
     +&\sum_{i=H,V}\left[ \ketbra{G}{X_i}\Omega_i(t) + \ketbra{X_i}{B}\Omega_i(t) \right] + \text{H.c.}~. \label{eqn:hamilton}
\end{align}
We incorporate electron-phonon coupling using the polaron approach described in \cite{bauch2021ultrafast,quilter2015phonon,glassl2013proposed,PhysRevB.105.045302}. Non-Markovian losses are considered using the Lindblad-operator 
\begin{align}
    \mathcal{L}_{\hat{O}}(\rho) = 2\hat{O}\rho\hat{O}^\dagger - \hat{O}^\dagger\hat{O}\rho - \rho\hat{O}^\dagger\hat{O}~. \label{eqn:lindblad}
\end{align}
for any given system operator $\hat{O}$. These include but are not limited to cavity losses ($\hat{O} = \sqrt{\kappa/2}\hat{b}_i$) and the radiative decay of the electronic state population ($\hat{O} = \sqrt{\gamma_\text{rad}/2}\ketbra{i}{j}, i\neq j$). 
The optical excitation pulse is considered by driving the electronic coherences using a Gaussian envelope \cite{bauch2021ultrafast,PhysRevB.105.045302}. Implementation details can be found at \cite{Bauch_QDLC-C_2023}.

\subsection{Purcell Factor}
Assuming full overlap between the cavity mode(s) and the electronic wave function of the QD states, the Purcell factor reads
\begin{align}
    \purcell &= \frac{\gamma_\text{Cavity}}{\gamma_\text{Rad}} \\
    &= \frac{3}{4\pi^2}\left(\frac{\lambda}{n}\right)^3\frac{\Q}{V_\text{mod}} \label{eqn:purcell_factor_em}\\
    &= \frac{2g^2}{\gamma_\text{Rad}\kappa}\frac{\kappa^2}{(\Delta E/\hbar)^2-\kappa^2}~. \label{eqn:purcell_factor_qd}
\end{align}
For the Maxwell simulation, $\lambda$ is the mode wavelength, $n$ is the refractive index of the cavity material, $\Q$ is the cavity $\Q$-factor and $V_\text{mod}$ the corresponding mode volume. Due to diverging fields in the calculation domain, it's challenging to estimate the later one effectively \cite{kristensen2012generalized}. Therefore, we don't calculate the Purcell enhancement through Eq. \eqref{eqn:purcell_factor_em}. Instead, we use the ratio of the power radiated by an electric point-dipole centered in the cavity compared to the power emitted in an infinite, homogeneous bulk material like described in \cite{Taflove:13}. This approach leads to the ratio of the local density of states (LDOS), which can be easily calculated by Fourier-transforming the fields neglecting the finite linewidth of the biexciton-exciton transition:
\begin{align} \mathcal{F}_{P,l}\left(\bm{x}_0, \omega \right) &= \frac{\mathrm{LDOS}_{l,\mathrm{cav}}\left(\bm{x}_0, \omega \right)}{\mathrm{LDOS}_{l,\mathrm{bulk}}\left(\bm{x}_0, \omega \right)}\\
	&= \frac{\mathrm{Re}\left(\mathcal{E}_{l, \mathrm{cav}}\left(\bm{x}_0, \omega \right) p(\omega)^{*} \right)}{\mathrm{Re}\left(\mathcal{E}_{l,\mathrm{bulk}}\left(\bm{x}_0, \omega \right) p(\omega)^{*} \right)},
 \end{align}
where $\mathcal{E}_l$ is the electric field at the point-dipole's position $\bm{x}_0$ and its corresponding index $l \in \lbrace x,y,z \rbrace$ defines the polarization, whereas $p(\omega)$ is the envelope of the point-dipole's current.

Translating the Purcell enhancement into parameters for the quantum dot simulation, the parameters then reduce to the light-matter coupling $\hbar g$, the radiative decay rate $\gamma_\text{Rad}$, the rate of cavity losses $\kappa$ and the cavity-exciton detuning $\Delta E = E_C - E_\XS$. A maximum Purcell enhancement $\purcell$ is achieved for a perfectly tuned cavity mode energy $E_c$ and electronic transition energy $E_\XS$. Substituting $\hbar\kappa=\frac{E_c}{\Q}$ into \cref{eqn:purcell_factor_qd} and assuming $E_c=E_\XS$, the QD-cavity coupling results in 
\begin{align}
    \hbar g = \sqrt{\frac{\purcell E_c}{2\Q}}\sqrt{\hbar \gamma_\text{Rad}}~. \label{eqn:coupling_em}
\end{align}
Only the front part of \cref{eqn:coupling_em} is approximated via Maxwell EM-field simulations. Because the radiative decay rate $\gamma_\text{Rad}$ is a property unique to every QD, it is assumed to be around $\hbar\gamma_\text{Rad} = \SI{2.5}{\mueV}$ in this work, which is a value well in the range of usual QD parameters.

\subsection{Indistinguishability and Visibility}
The two-time correlation functions $G^{(i)}(t,\tau)$ are evaluated using the quantum regression theorem, which is an approximation suitable for low photon numbers and for the typical GaAs QD parameters used in this work \cite{cosacchi2021accuracy}. They are defined as
\begin{align}
    G^{(1)}(t,\tau) &= \expval{ \hat{a}_i^\dagger(t+\tau)\hat{a}_j(t) }~, \label{eqn:g1}\\
    G^{(2)}(t,\tau) &= \expval{ \hat{a}_i^\dagger(t)\hat{a}_j^\dagger(t+\tau)\hat{a}_k(t+\tau)\hat{a}_l(t) }~\label{eqn:g2}.
\end{align}

For the integrated two-time second order correlation function 
\begin{align}
\Bar{\Bar{G}}^{(2)}=\int \int G^{(2)}(t,\tau)\d\tau\d t\approx 0~,\label{eqn:double_time_integrated_g2}
\end{align}
the visibility

\begin{align}
    \mathcal{V} = \frac{2\int_0^{\mathcal{T}}\int_0^{\mathcal{T}-t} |G^{(1)}(t,\tau)|^2 \d\tau\d t }{\left(\int_0^{\mathcal{T}}G^{(1)}(t,0)\d t\right)^2} \label{eqn:visibility}
\end{align}
becomes a sufficient figure of merit for the QD emission and thus for the single photon indistinguishability \cite{scholl2020crux,fischer2016dynamical}.

Because \cref{eqn:double_time_integrated_g2} is not generally true, especially when electron-phonon interactions generate two-photon states, we define the single-photon HOM-indistinguishability \cite{PhysRevB.98.045309,bauch2021ultrafast} at time $\mathcal{T}$ as
\begin{widetext}
\begin{align}
    \indist(\mathcal{T}) = 1-p_c = 1-\frac{\int_0^{\mathcal{T}}\int_0^{{\mathcal{T}}-t} \left( \ev{\hat{a}^\dagger\hat{a}}(t)\ev{\hat{a}^\dagger\hat{a}}(t+\tau) + G^{(2)}(t,\tau) - \abs{G^{(1)}(t,\tau)}^2 \right)\d \tau\d t}{\int_0^{\mathcal{T}}\int_0^{\mathcal{T}-t}\left( 2\ev{\hat{a}^\dagger\hat{a}}(t)\ev{\hat{a}^\dagger\hat{a}}(t+\tau)- \abs{\ev{\hat{a}(t+\tau)} \ev{\hat{a}^\dagger(t)}}^2 \right)\d \tau\d t}~. \label{eqn:homindist}
\end{align} 
\end{widetext}
Here, $p_c$ is the coincidences-count, and $G^{(1)}(t,\tau) = \ev{\hat{a}^\dagger(t+\tau)\hat{a}(t)}$ and $G^{(2)}(t,\tau) = \ev{\hat{a}^\dagger(t)\hat{a}^\dagger(t+\tau)\hat{a}(t+\tau)\hat{a}(t)}$ are the first and second order two-time correlation functions \cite{bauch2021ultrafast}, respectively. The quantum regression theorem \cite{cosacchi2021accuracy} is used to approximate these values.

While not generally true, for the results shown in this work, a high indistinguishability will usually also result in a high visibility.

To include the maximum degree of coherence possible when approximating the degree of polarization entanglement from the emitted modes, we use $\Bar{\Bar{G}}^{(2)}$, as discussed in \cite{bauch2021ultrafast,cygorek2018comparison}. 

Because we require low $\Q$-factors for our simulations, we do not separate the emission from the radiative decay of the QD states and the emission from the optical resonator. Instead, we compose a super operator
\begin{align}
\hat{a}_\XB^\dagger &= \ketbra{\XB}{\XS_i} + \hat{b}^\dagger_i~~,\hat{a}_\XS^\dagger = \ketbra{\XS_i}{\XG}~,
\end{align}
with the electronic polarization mode $i\in\left\{H,V\right\}$. This is also the reason why in \cref{fig:poc_biexciton} the concurrence is reduced to near zero for a cavity at the two-photon resonance, while the indistinguishability increases, which is uncommon for ordinary biexciton-cavity emissions \cite{heinze2017polarization}. Note, that for a very broad cavity, where the cavity line width captures both transitions, the super operator experiences additional losses due to the mixing of biexciton-exciton and exciton-ground transitions. This means the indistinguishability calculated should be lower than the experimentally measured values due to the spectral filtering of the detectors, which is usually missing in the simulations.

\subsection{Optimization Details}\label{subsec:OptDetails}
For our Maxwell optimizations, we considered the following six degrees of freedom for the CBG device: grating period, first trench width, remaining trench width, cavity radius, height of the CBG and the thickness of the SiO\textsubscript{2} layer. The thickness of the gold mirror was fixed to \SI{100}{\nano \meter}, since its high conductivity at \SI{4.2}{\kelvin} leads to a small skin depth of the fields. As refractive index for the InGaAs-CBG we assumed $n_\text{InGaAs} = 3.4$ at \SI{4.2}{\kelvin} \cite{bremer2022numerical}.

For the batches of the global optimization we used the MPI-subgroup feature of Meep to assign each subgroup to one AMD Milan 7763 CPU with 64 cores. This scheme avoids costly communication between sockets and computation nodes in a cluster environment. Therefore, the parallelization is only limited by the maximum batch size of the optimization method, which is roughly twenty \cite{gonzalez2016}. Furthermore, we used the expected improvement acquisition function for ten batches with 1k-2k iterations, so 10k-20k goal function evaluations in total, which were mostly sufficient for finding an appropriate global optimum. One iteration took about \SI{1}{\minute} \SI{30}{\second} in mean for the low-$\Q$ approach and about \SI{3}{\minute} \SI{30}{\second} for maximum Purcell enhancement due to the fixed energy criterion of \SI{-40}{\decibel}.

For the 3D-optimization in CST-MWS we used two NVIDIA A100 GPUs in parallel and achieved a runtime of just over \SI{20}{\minute} per iteration with a resolution of at least 40 computational cells per wavelength. About 400-500 iterations were required for a sufficient convergence of the Nelder-Mead optimizer.

\begin{figure}[b]
	\includegraphics[width=\columnwidth]{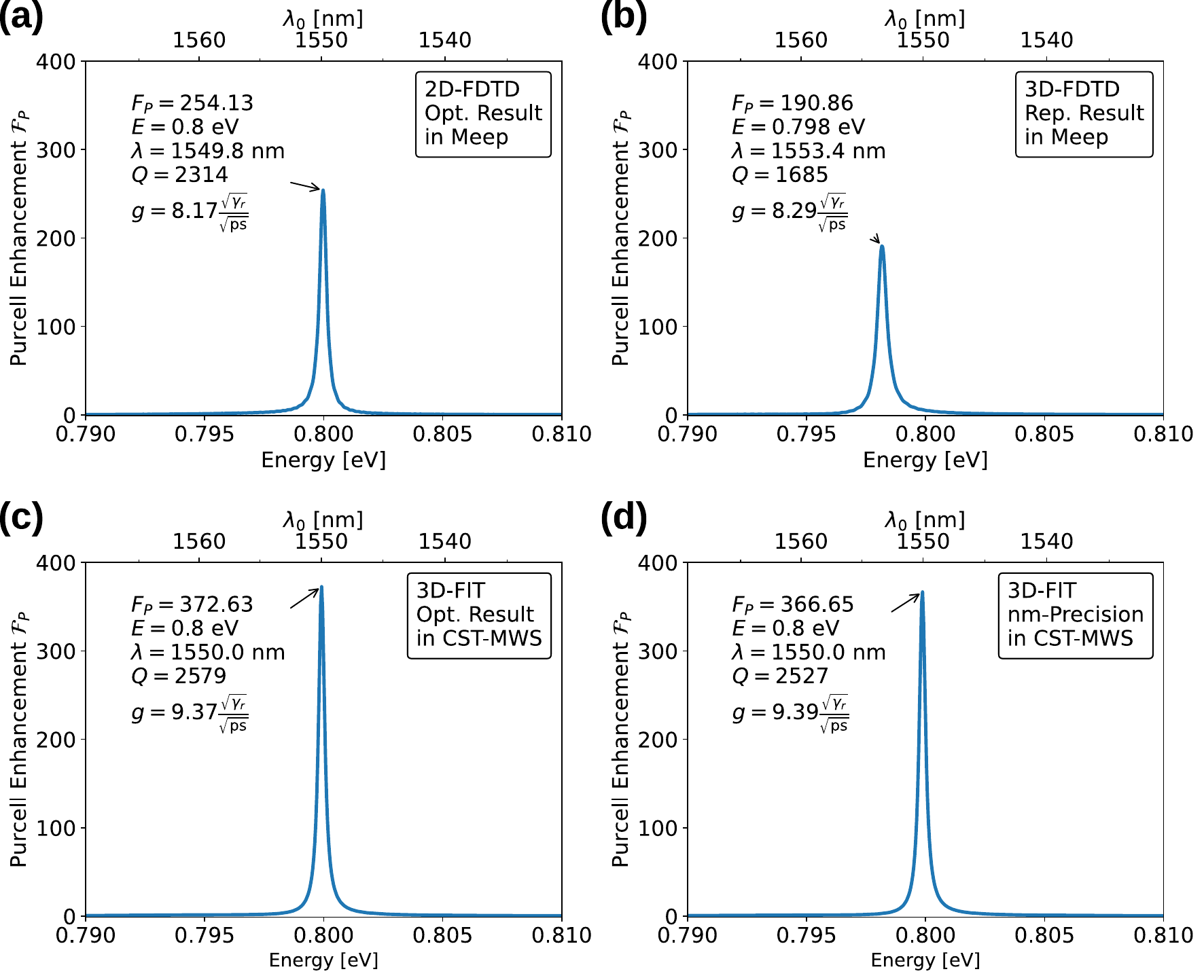}
	\caption{\label{fig:Plots_maxFP} \textbf{(a)} Pre-optimized 2D-FDTD result in Meep. \textbf{(b}) Reproduction of the former 2D-FDTD optimum for a 3D-FDTD simulation in Meep. \textbf{(c)} Polished 3D-FIT optimization result in CST-MWS. \textbf{(d)} Reproduction of the former 3D-FIT optimum with rounded dimensions to nanometer precision.}
\end{figure}

\subsection{Maxwell Optimization for Maximum Purcell Factor}\label{subsec:maxFP}
Our first trial to optimize the CBG resonator for a maximum Purcell factor led to impressive results showing a Purcell enhancement of more than 370. The cost function simply benefits as high as possible Purcell factors while punishing deviations from the target wavelength and upcoming unwanted modes. The latter was realized by an added spectrum integral excluding the region around the target wavelength for a Lorentzian shaped mode with an assumed $\Q$-factor of e.g. 700. The results for Meep and CST-MWS are shown in \cref{fig:Plots_maxFP}. \cref{fig:Plots_maxFP},b shows the 3D-FDTD reproduction in Meep from the optimization result of the 2D-FDTD optimization in \cref{fig:Plots_maxFP},a. The deviations in the Purcell enhancement and target wavelength reveal, that it is not sufficient only running 2D-simulations to obtain accurate results for a fabrication process. Instead, doing optimizations on GPU accelerators afterwards compensates those deviations shown in \cref{fig:Plots_maxFP},c. Here, we could even improve the former 2D-FDTD result by starting a few optimization runs adjusting the parameter bounds until we reached a Purcell factor of more than 370. Since the optical properties of the structure benefit from dimensions in the picometer range, which are purely synthetical, we provide another result with rounded dimensions to nanometer precision in \cref{fig:Plots_maxFP},d. 

One major key for finding these CBG resonators exhibiting such a high Purcell enhancement was the additionally added first trench width of the CBG as another degree of freedom, which was neglected by all recent publications investigating CBG resonators for optical properties. The first trench width of the result in \cref{fig:Plots_maxFP},d is significantly larger (\SI{204}{\nano \meter}) compared to the remaining ones (\SI{93}{\nano \meter}).

\subsection{Temperature Dependency of the Linear Optimization}\label{subsec:tempdep}
\begin{figure}[h!]
\includegraphics[width=\columnwidth]{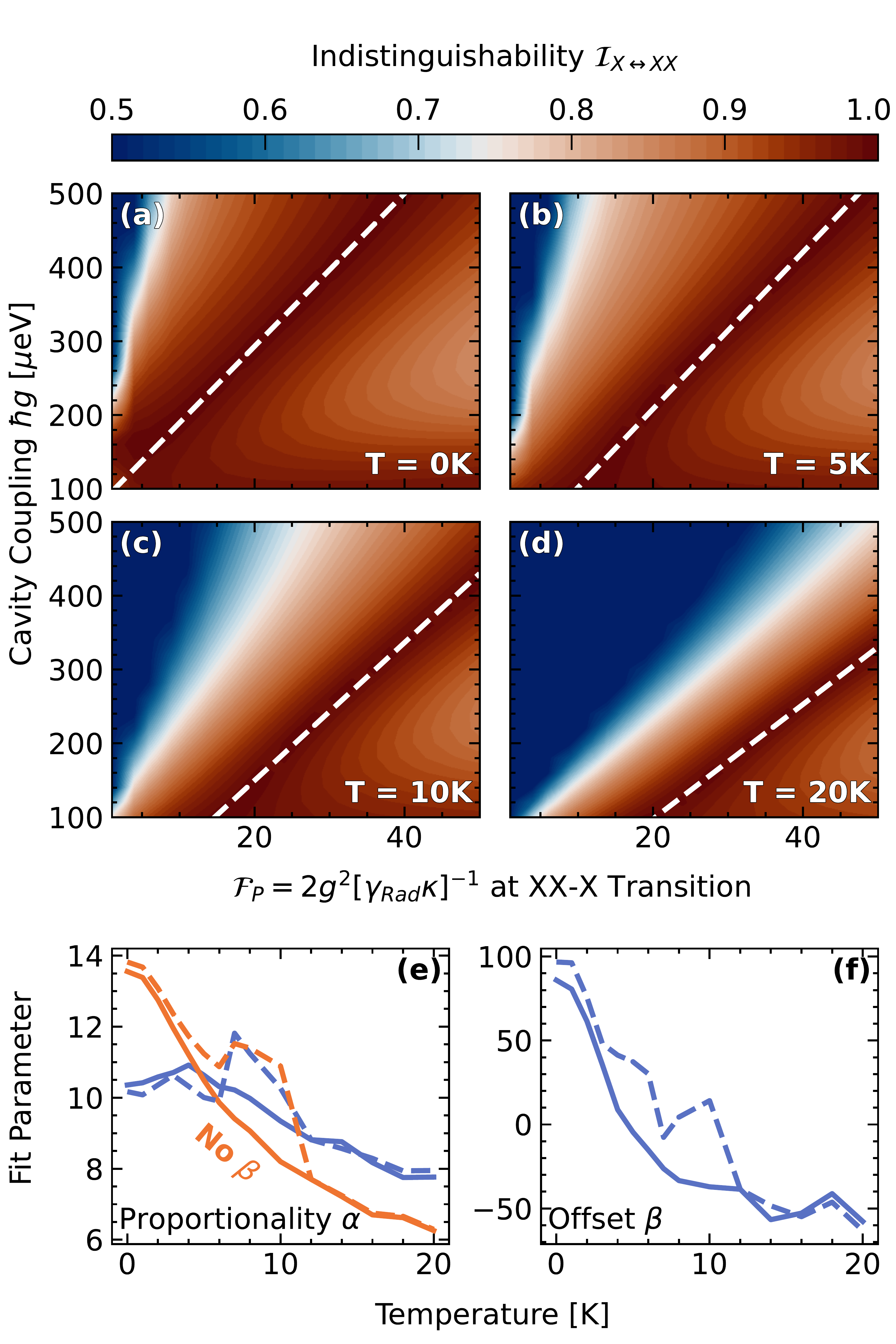}
\caption{\label{fig:tempdep}Temperature dependency of the fitting parameters $\alpha(T)$ and $\beta(T)$. Four different temperatures are shown exemplary for $T\in{0,5,10,20}$. The settings for the simulations correspond to \cref{fig:g_kappa_sweepT4.2k}, including the linear fit (white dashed lines). The temperature dependency for the fitting parameters is shown for the proportionality $\alpha(T)$ (e) and the offset $\beta(T)$ (f). Fitting with a nonzero offset (violet lines) are compared with a fit through the origin (orange lines). Fitting only the biexciton-exciton indistinguishability $\indist_\XB$ (solid lines) is compared to fitting the product of all indistinguishabilities and the concurrence (dashed lines).}
\end{figure}

Varying the temperature for the configuration given in \cref{fig:g_kappa_sweepT4.2k} reveals the fitting parameters with scaling $\alpha(T)$ and offset $\beta(T)$, which is shown in \cref{fig:tempdep}. The reciprocal scaling of the fitting parameters may be induced by the also reciprocal temperature scaling of the phonon interactions. We see a slight decline in the scaling $\alpha$ with raising temperature, and when disabling the linear offset $\beta$, a pure reciprocal scaling can be observed. 
Because these simulations are complex and very time-intensive, we calculate the temperature dependency of the fitting parameters at low-resolution (\cref{fig:tempdep},e,f) and hence exclude them from the main text. Fitting the biexciton-exciton photons indistinguishability does not significantly differ from fitting the product of all three quantum properties. Nevertheless, fitting the product can lead to slight artifacts, as can be seen in \cref{fig:tempdep},e,f.

\subsection{CBG Dimensions}\label{subsec:cbgdim}
\begin{table}[h!]
	\renewcommand{\arraystretch}{1.2}
	\centering
	\begin{tabular}{ |c||c|c| }
		\hline 
		\textbf{Dimension [nm]} & \textbf{Fig. \ref{fig:maxwell_optimized_cavities},a} & \textbf{Fig. \ref{fig:Plots_maxFP},c} \\ [0.5ex]
		\hline
		\hline
		Grating Period & 563.3 & 561.9\\
		\hline
		1. Trench Width & 154.3 & 204.5\\
		\hline
		Trench Width & 161 & 93.3\\
		\hline
		Cavity Radius & 584.1& 588.1\\
		\hline
		CBG Height & 450.6& 363.8 \\
		\hline
		Thickness SiO\textsubscript{2} & 359.6 & 189.6\\
		\hline
	\end{tabular}
\end{table}

\clearpage
\bibliography{bibo}

\end{document}